\documentclass[fleqn,usenatbib]{mnras}

\usepackage{newtxtext,newtxmath}
\usepackage{hyperref}
\usepackage[T1]{fontenc}

\DeclareRobustCommand{\VAN}[3]{#2}
\let\VANthebibliography\thebibliography
\def\thebibliography{\DeclareRobustCommand{\VAN}[3]{##3}\VANthebibliography}

\usepackage{graphicx}	% Including figure files
\usepackage{amsmath}	% Advanced maths commands
\usepackage{tabularx}
\title[\textit{Swift}-BAT blazars with \textit{NICER}]{Investigating the variability of \textit{Swift}-BAT blazars with \textit{NICER}}

\author[Mundo \& Mushotzky]{
Sergio A. Mundo$^{1}$\thanks{E-mail: smundo@astro.umd.edu}
and Richard Mushotzky
\\
$^{1}$Department of Astronomy, University of Maryland, College Park, MD 20742, USA\\
}

\date{Accepted XXX. Received YYY; in original form ZZZ}

\pubyear{2022}

\begin{document}
\label{firstpage}
\pagerange{\pageref{firstpage}--\pageref{lastpage}}
\maketitle

\begin{abstract}
We present results of X-ray spectral and time-domain variability analyses of 4 faint, ``quiescent” blazars from the \textit{Swift}-BAT 105-month catalog. We use observations from a recent, 5-month long \textit{NICER} campaign, as well as archival BAT data. Variations in the 0.3-2 keV flux are detected on minute, $\sim$weekly, and monthly timescales, but we find that the fractional variability $F_{\rm var}$ on these timescales is $<$25\% and decreases on longer timescales, implying generally low-amplitude variability across all sources and showing very low variability on monthly timescales ($F_{\rm var}\lesssim13\%$), which is at odds with previous studies that show that blazars are highly variable in the X-rays on a wide range of timescales. Moreover, we find that the flux variability on very short timescales appears to be characterized by long periods of relative quiescence accompanied by occasional short bursts, against the relatively time-stationary nature of the variability of most other AGN light curves. Our analysis also shows that the broadband X-ray spectra (0.3-195 keV) of our sources can be described with different power law models. As is the case with most blazars, we find that 2 sources (2MASS J09343014-1721215 and PKS 0312-770) are well-modeled with a simple power law, while the remaining two (1RXS J225146.9-320614 and PKS 2126-15) exhibit curvature in the form of a log-parabolic power law. We also find that, in addition to the continuum, PKS 2126-15 requires significant absorption at the soft X-rays ($\lesssim$1 keV) to fully describe the observed curvature, possibly due to absorption from the intergalactic medium.
\end{abstract}


%%%%%%%%%%%%%%%%%%%%%%%%%%%%%%%%%%%%%%%%%%%%%%%%%%

%%%%%%%%%%%%%%%%% BODY OF PAPER %%%%%%%%%%%%%%%%%%

\section{Introduction}

Blazars are a class of radio-loud (RL) active galactic nuclei (AGN) whose defining characteristic is a jet aligned close to the observer’s line of sight (\citealt{1995PASP..107..803U}). Due to the bulk relativistic motion of the particles in the jet, any emission from the latter will be relativistically beamed in the direction of motion, and an observer at rest will detect emission that is much more powerful than if the particles were at rest. As a result, the emitted radiation from blazars, which is dominated by extreme non-thermal processes that take place in the jet, has historically been known to be very luminous, with high-amplitude, rapid variations in flux, spectra, and polarization observed across most timescales and energy bands \citep[see e.g.][]{1976ARA&A..14..173S,1978PhyS...17..265B,1980ARA&A..18..321A,1985ApJ...298..114M,1986ApJ...306L..71M,1986ApJ...302..337F,1995ARA&A..33..163W,1997ARA&A..35..445U,2005A&A...442...97A,2008A&A...486..721L,2014A&A...563A..57S}.

While most blazars share a number of characteristics, such as flat radio spectra, rapid variations in flux and in radio/optical polarization, and superluminal motion at radio wavelengths \citep[]{1990ApJ...352...81M,1994ApJ...430..467V,2005AJ....130.1418J}, they can nonetheless be separated into two subclasses: BL Lacertae (BL Lac) objects and flat-spectrum radio quasars (FSRQs). This classification is usually based on the rest-frame equivalent width (EW) of the optical emission lines, with FSRQs showing broad lines with EW$>$ 5\AA \ and BL Lacs showing weak or no emission lines in their spectra \citep[]{1991ApJ...374..431S}. 

FSRQs and BL Lacs can further be told apart by their broadband spectral energy distributions (SEDs). The SEDs of BL Lacs are characterized by two broad emission humps that are usually close to equally luminous. The peak at low frequencies is likely caused by synchrotron emission processes that usually emit from the infrared to the X-rays, while the peak at high frequencies (X-rays to $\gamma$-rays) likely arises from the synchrotron self-Compton (SSC) mechanism, an inverse Compton process by the same highly energetic particles in the jet. FSRQs show the same two aforementioned emission humps, but in these blazars, the Compton hump is more luminous than the synchrotron hump; this is likely due to the radiatively efficient accretion in the more luminous FSRQs, which leads to a UV-bright disk that produces ionizing photons to form a broad-line region (BLR). The BLR, in turn, provides an additional source of photons external to the jet that undergo inverse Compton scattering \citep[]{2009MNRAS.396L.105G,2011MNRAS.414.2674G}, producing the high-energy hump in the SEDs of FSRQs. 

%This external Compton (EC) mechanism can also occur with external photon fields other than the one from the BLR, such as seed photons from the disk itself and infrared emission from the torus (see Ghisellini \& Tavecchio 2009 for relevant modeling).

%as well as occasionally a third hump in between, which arises from thermal emission from the accretion disk.

In particular, the X-ray spectra of blazars can generally be well-described by a simple power law or a curved continuum that represent the non-thermal processes occurring in the jet \citep[see e.g.][]{1997ApJ...480..534C}. Many spectra show curvature in the form of either a log-parabolic power law or a broken power law \citep[e.g.][]{2004A&A...413..489M,2006A&A...448..861M,2009A&A...504..821P,2013ApJ...770..109F,2018A&A...616A.170A,2021MNRAS.508.1701D}, with many blazar emission models predicting that spectra exhibiting either shape can result from a relativistic particle distribution that has a similar curvature (see e.g. \citealt{1994ApJ...421..153S,1997ApJ...484..108S,2009ApJ...704...38S,2007ApJ...665..980T,2008MNRAS.386..945T,2009MNRAS.397..985G,2015MNRAS.448.1060G,2018A&A...616A.170A}, and references therein). Alternatively, curvature might be observed simply due to the location of the X-rays on the SED, as is the case with ``extreme" blazars that have the peak of their synchrotron hump located at very high frequencies ($\nu_{syn}^{peak}\gtrsim10^{17}$ Hz; see e.g. \citealt{2019ApJ...881..154P} for examples), likely due to the efficient acceleration mechanisms in their jets \citep[e.g.][]{2001A&A...371..512C,2019MNRAS.486.1741F,2019ApJ...882L...3P}.

%There has, however, been some debate on whether the curvature observed in the emission spectra is completely intrinsic to the source and its host galaxy, or whether the culprit is external in nature.

%(see e.g. Massaro et al. 2004, 2006; Arcodia et al. 2018; Dalton et al. 2021).

%More specifically, a log-parabolic curved X-ray spectrum could arise from a log-parabolic particle distribution, or a simple power law particle distribution with a cooled high-energy tail (Paggi et al. 2009; Furniss et al. 2013). 

%In addition, blazar emission models can predict intrinsic spectral breaks in the emitted spectra that could result from breaks in the injected particle spectrum, which could in turn be caused by inefficient cooling of low-energy electrons (see e.g. Sikora et al. 1994, 1997, 2009; Tavecchio et al. 2007; Tavecchio \& Ghisellini 2008; Ghisellini \& Tavecchio 2009, 2015; Arcodia et al. 2018, and references therein), leading to broken power law X-ray spectra. 

An intrinsic curved continuum does not always yield a complete picture of the X-ray emission from a blazar. Certain studies have shown that sometimes, significant photoelectric absorption, in addition to said continuum, is required in the soft X-rays to fully describe the curved X-ray spectrum \citep[e.g.][]{1997ApJ...478..492C,2000ApJ...543..535T,2001MNRAS.323..373F,2001MNRAS.324..628F,2004MNRAS.350..207W,2004MNRAS.350L..67W,2006MNRAS.368..844W,2005MNRAS.364..195P,2004AJ....127....1G,2006AJ....131...55G,2007ApJ...669..884S,2013ApJ...774...29E,2018A&A...616A.170A,2021MNRAS.508.1701D}. At first glance, this contradicts the very nature of blazars, since in general these objects are considered to have negligible X-ray absorption along the line of sight, due to their kiloparsec-scale jet likely sweeping away any potential contribution to absorption from the host galaxy. In order to reconcile this, several studies \citep[e.g.][]{2001MNRAS.323..373F,2011ApJ...734...26B,2012MNRAS.421.1697C,2015A&A...575A..43C,2013MNRAS.431.3159S,2013ApJ...774...29E,2018A&A...616A.170A,2021MNRAS.508.1701D} have suggested that the absorption might be due to the highly ionized ``warm-hot" intergalactic medium (WHIM), with more recent studies emphasizing that such a component from the intergalactic medium should be considered in the spectral analysis of blazars when appropriate \citep[see e.g.][]{2018A&A...616A.170A,2021MNRAS.508.1701D}.

%possibly omit paragraph above

%Since FSRQs are among the most powerful blazars, they are ideal candidates to probe a possible absorption contribution from the IGM.

In a preliminary time-domain variability analysis of 117 blazars in the \textit{Swift}-Burst Alert Telescope (BAT) 105-month catalog (Mundo et al. 2022, in preparation), we find that a non-negligible fraction ($\sim$30\%) of the sample does not show statistically significant variability on monthly timescales in the 14-195 keV band, in tension with previous works that have established that blazars are extremely variable objects at almost every timescale and wavelength. However, it is unclear from just the BAT data if this apparent lack of variability is a result of truly relatively constant emission (i.e. only moderate amplitude variability) or emission that is mostly constant with occasional flaring events, as seen in Fermi data for some sources \citep[e.g.][]{2015ApJ...803...15P,2015ApJ...807...79H}, or if it is solely related to the sensitivity or systematic issues of the BAT data. The BAT catalog data are unique in providing continuous observations over a 9-year timescale for a hard X-ray selected sample and thus sample the time variability of these objects in a previously unexamined time domain. Therefore, confirmation of these results could change our understanding of the properties of blazars and beamed AGN.

%maybe change "or emission that is mostly constant..." to just a comma instead of the or

%The BAT’s relatively low sensitivity per unit time (Tueller et al. 2008) means that it can only constrain the variability on monthly timescales for the vast majority of objects detected in the 105-month catalog. 

Because of the BAT's relatively low sensitivity per unit time \citep{2008ApJ...681..113T}, it can only constrain the variability on monthly timescales for the vast majority of objects detected in the 105-month catalog. In order to determine if the supposedly ``non-varying" sources exhibit variability on shorter timescales not detected by the BAT, we started a $\sim$5-month long campaign in 2021 with the Neutron Star Interior Composition Explorer (\textit{NICER}) for each of 4 such apparently non-variable sources. Thanks to the \textit{NICER} telescope's $>$100 times sensitivity per unit time compared to the BAT, the campaign allows for an estimate of the variability on a wide range of timescales, probing shorter timescales while also representing the timescales of the BAT catalog.

%of eight 500s pointings per month, for 5 months per source,
% (see Table 1)

\begin{table*}
	\centering
	\caption{\textit{NICER} observations and sources used in our analysis. Shown are the source name, blazar type, redshift, observation ID, observation start dates, and the exposure times. \textit{1RXS J225146.9-320614}: 38 observations with exposures between $\sim$200s and $\sim$2 ks; \textit{2MASS J09343014-1721215}: 49 observations with exposures between $\sim$200s and $\sim$3 ks; \textit{PKS 2126-15}: 41 observations with exposures between $\sim$150s and $\sim$4 ks; \textit{PKS 0312-770}: 35 observations with exposures between $\sim$100s and $\sim$2 ks. For all sources, observations are spaced across 5 months. The wide range of exposures in the observations is the result of flagging periods of high background according to the \texttt{nicerl2} pipeline criteria. The full list of observations can be found in the electronic version and in the HEASARC archive.}
	\label{tab:obs}
	\begin{tabularx}{\textwidth}{cccccr} % four columns, alignment for each
		\hline
		Source name & Blazar Type & $z$ & Obs. ID & Start Date & Exposure (s)\\
		\hline
		1RXS J225146.9-320614 & BL Lac & 0.2460 & 4638020101 & 2021/06/20 & 663\\
		 &  &  & 4638020201 & 2021/06/23 & 1247\\
		 &  &  & 4638020301 & 2021/06/27 & 1268\\
		 &  &  & 4638020401 & 2021/06/29 & 653\\
		 &  &  & ... & ... &...\\
		2MASS J09343014-1721215 & BL Lac & 0.2499 & 4638030101 & 2021/03/03 &582\\
		 &  &  & 4638030201 & 2021/03/06 &687\\
		 &  &  & 4638030301 & 2021/03/10 &1192\\
		 &  &  & 4638030501 & 2021/03/17 &675\\
		 &  &  & ... & ... &...\\
		PKS 2126-15 & FSRQ & 3.2680 & 4638010201 & 2021/06/22 &710\\
		 &  &  & 4638010301 & 2021/06/26 &1480\\
		 &  &  & 4638010401  & 2021/06/29 &4284\\
		 &  &  & 4638010402 & 2021/06/30 &3405\\
		 &  &  & ... & ... &...\\
		PKS 0312-770 & FSRQ & 0.2230 & 4638040101 & 2021/04/10 & 404\\
		 &  &  & 4638041301 & 2021/04/13 & 1234\\
		 &  &  & 4638040301 & 2021/04/17 & 364\\
		 &  &  & 4638040401 & 2021/04/20 & 89\\
		 &  &  & ... & ... &...\\
		\hline
	\end{tabularx}
\end{table*}

The 4 sources analyzed here with new \textit{NICER} data (2MASS J09343014-1721215; PKS 0312-770; 1RXS J225146.9-320614; PKS 2126-15) correspond to the brightest blazars from the \textit{Swift}-BAT 105-month catalog for which there was very little variability ($F_{\rm var}\lesssim10\%$) on monthly timescales. In general, these objects lie at the low flux end of the BAT blazar population and are among the most quiescent sources in the catalog. In addition, according to a recent spectroscopic study of the blazars in the 105-month catalog \citep{2019ApJ...881..154P}, the \textit{NICER} band (0.3-10 keV) falls on the same hump as the BAT band (14-195 keV) for each of the 4 sources, meaning that the broadband X-rays are likely produced by the same underlying process for each source, and thus the new \textit{NICER} observations would shed light on the physical processes driving the BAT band. 

In this paper, our main objectives are to determine whether these 4 sources are variable, as well as to characterize the nature of their variability and their spectra in the \textit{NICER} band. We describe the observations and data reduction in Section 2, present our variability and spectral analyses and their results in Section 3, and discuss these results in Section 4.

\section{Observations and Data Reduction} \label{sec:obs}

\subsection{\textit{NICER}}

Each of the 4 sources in this investigation were observed over a period of 5 months in order to mimic the long timescales of the BAT 105-month catalog, with $\sim$8 observations per month that had varying individual exposure ($\sim$40 observations total per source, PI: Mundo; see Table \ref{tab:obs} and caption). Each source was observed for a total exposure of at least $\sim$20 ks taken over this 5-month period (all the data are available on the electronic version and the HEASARC archive\footnote{https://heasarc.gsfc.nasa.gov/cgi-bin/W3Browse/w3browse.pl}). The \textit{NICER} data were reduced using HEASOFT v.6.29.1 and the current calibration files available (v. xti20210707). The \texttt{nicerl2} pipeline was used with the default settings using all 56 detectors and applying the necessary filters and calibration to produce cleaned events files.

We used the tool \texttt{xselect} to extract 1-second binned light curves, which we later re-binned to three different timescales for our time-domain variability analysis (see Figure \ref{fig:pks2126lc} for an example of the campaign with light curves on minute and $\sim$weekly timescales for one of our sources). The background estimation and calculation of the total spectra were performed using the tool \texttt{nibackgen3C50}\footnote{https://heasarc.gsfc.nasa.gov/docs/nicer/tools/nicer\_bkg\_ \ est\_tools.html} \citep{2022AJ....163..130R}. The spectra were binned to a minimum of 20 counts per bin using the \texttt{grppha} tool.

\begin{figure}
	\includegraphics[width=\columnwidth]{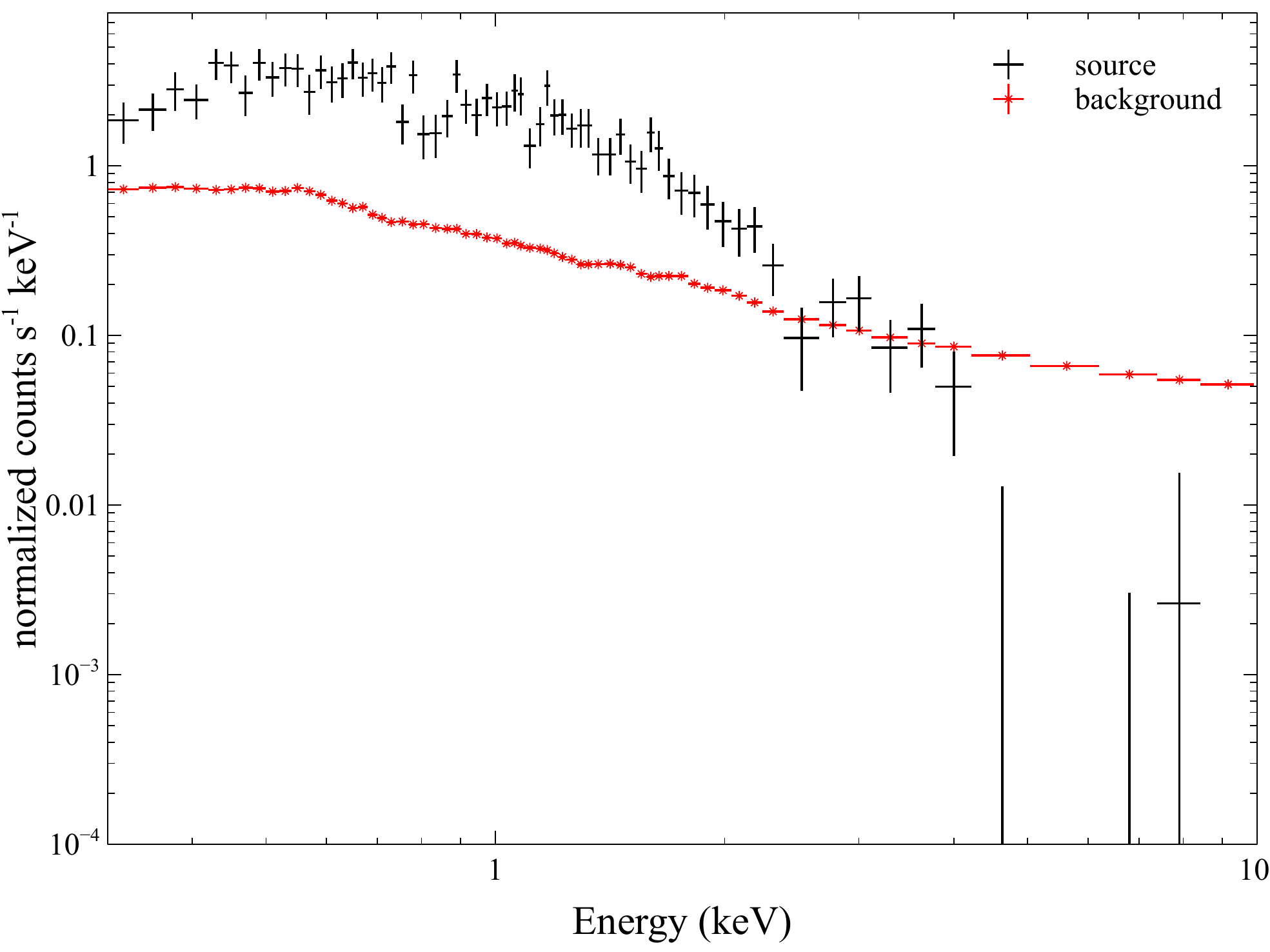}
    \caption{Sample spectrum of an individual observation of 2MASS J09343014-1721215 which took place during the second month of the campaign. The background begins to reach the level of the signal between 2-3 keV ($\sim2$keV). For consistency across all sources and observations, light curves are extracted in the 0.3-2 keV range for the flux variability analysis.}
    \label{fig:2masssampspec}
\end{figure}

\subsection{\textit{Swift}-BAT}

To probe the hard X-rays and complete the broadband X-ray spectrum, we make use of the archival observations available in the \textit{Swift}-BAT 105-month catalog\footnote{https://swift.gsfc.nasa.gov/results/bs105mon/} from the BAT Hard X-ray Survey \citep{2018ApJS..235....4O}. We include the ready-to-use, eight-channel time-averaged spectra that are provided in our analysis.

\begin{figure*}
    \centering
	\includegraphics[width=\textwidth]{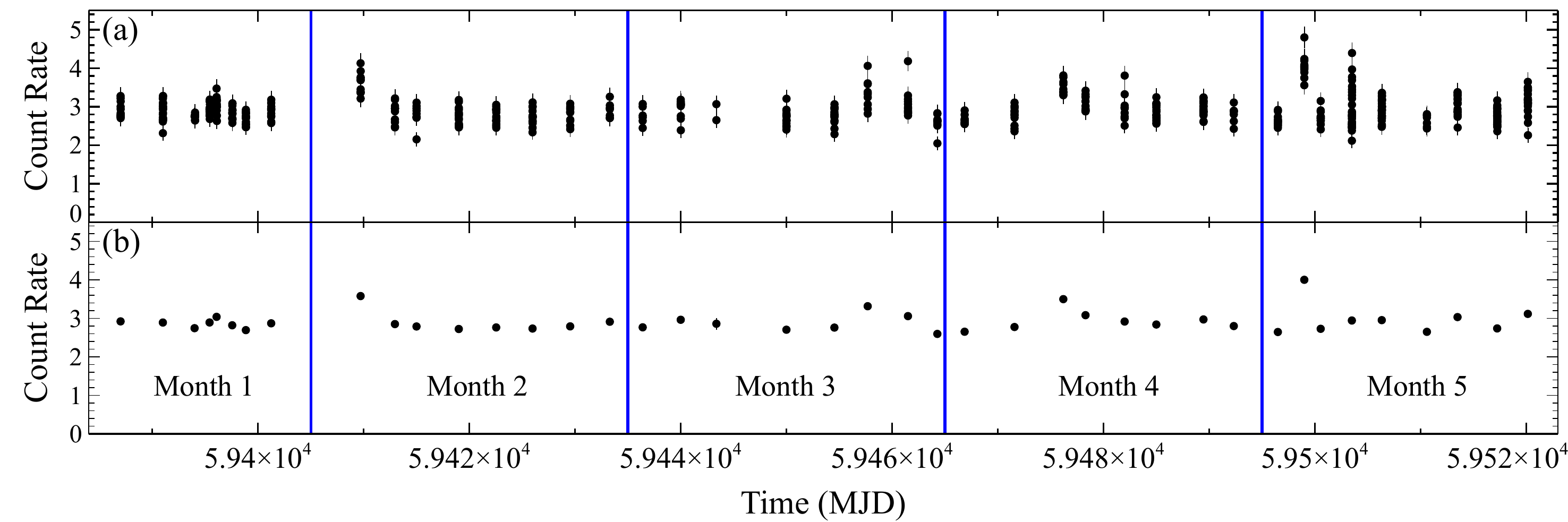}
	\includegraphics[width=\textwidth]{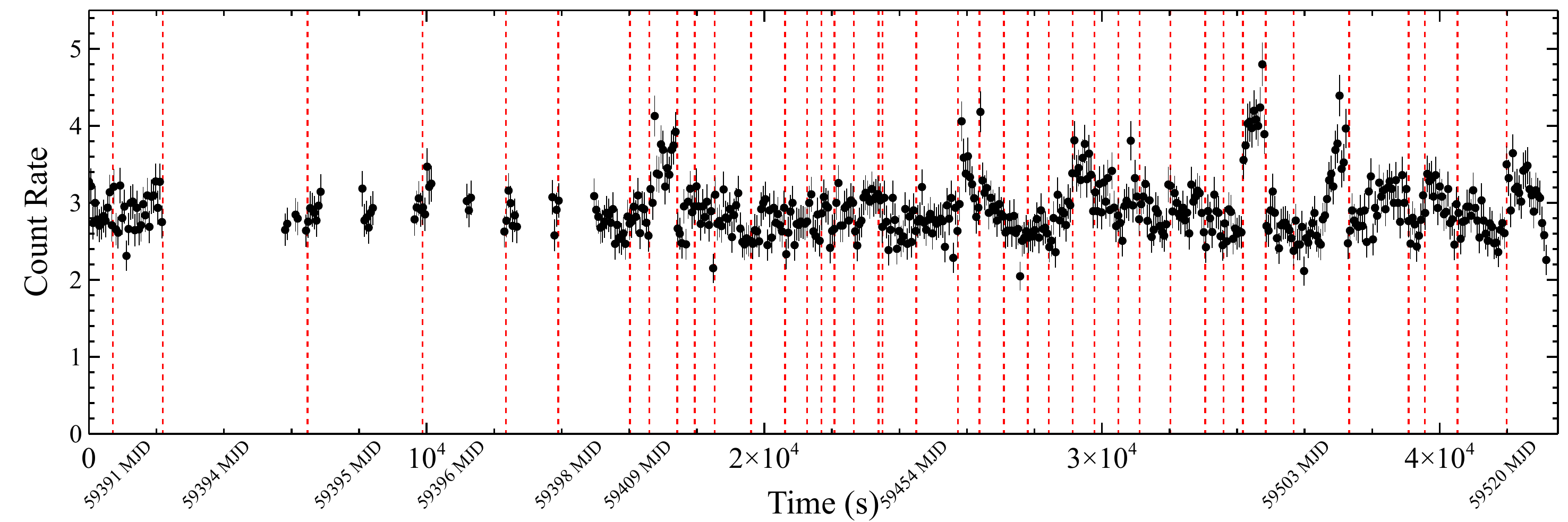}
    \caption{\textit{Top panel}: Sample light curve for one of our sources, PKS 2126-15, in two timescales probed in this study: (a) binned to one minute, and (b) binned $\sim$weekly (by observation). Count rates are in the 0.3-2 keV band. Blue lines are shown to depict the observational campaign structure. \textit{Bottom panel}: The same as top panel (a), but this time plotted over the total exposure time for the source in seconds, as opposed to over the total campaign time, in order to highlight the short timescale variability. Red dashed lines indicate a change from one observation to the next. To connect the bottom panel to top panel (a), the rate at an effective time of e.g.$\sim$1.7$\times$10$^{4}$ sec in the bottom panel is seen in Month 2 of top panel (a) at 59409.7 MJD. We also include labels on the bottom panel to show when certain observations occur in MJD. As with top panel (a), the data points are $\sim$60 s wide.  \textit{(Similar light curves for the other blazars are shown in Figures 1d-1i in the electronic version.)}}
    \label{fig:pks2126lc}
\end{figure*}

\section{Results} \label{sec:results}

\subsection{Time-domain flux variability analysis in the \textit{NICER} band}

We begin by performing a time-domain analysis of the flux variability for each source on minute, $\sim$weekly, and monthly timescales, which cover the range of timescales represented by both \textit{NICER} and the BAT. Since \textit{NICER} is not an imaging instrument, we need to rely on the spectrum and the estimated background from each observation to determine which band to extract the light curves from. We find that across all observations, the ratio of the signal to the background reaches $\sim$1 at the very least at $\sim$2 keV (see Figure \ref{fig:2masssampspec} for an example), with some observations showing the background reaching the signal above that energy. In order to have a consistent comparison across all observations and sources, we therefore decide to extract light curves in the 0.3-2 keV range for each source (see e.g. Figure \ref{fig:pks2126lc}) and perform our flux variability analysis in this range. (We show the light curves to the additional 3 sources in Figures 1d-1i in the electronic version.)

%signal is above the background

\begin{figure}
	\includegraphics[width=\columnwidth]{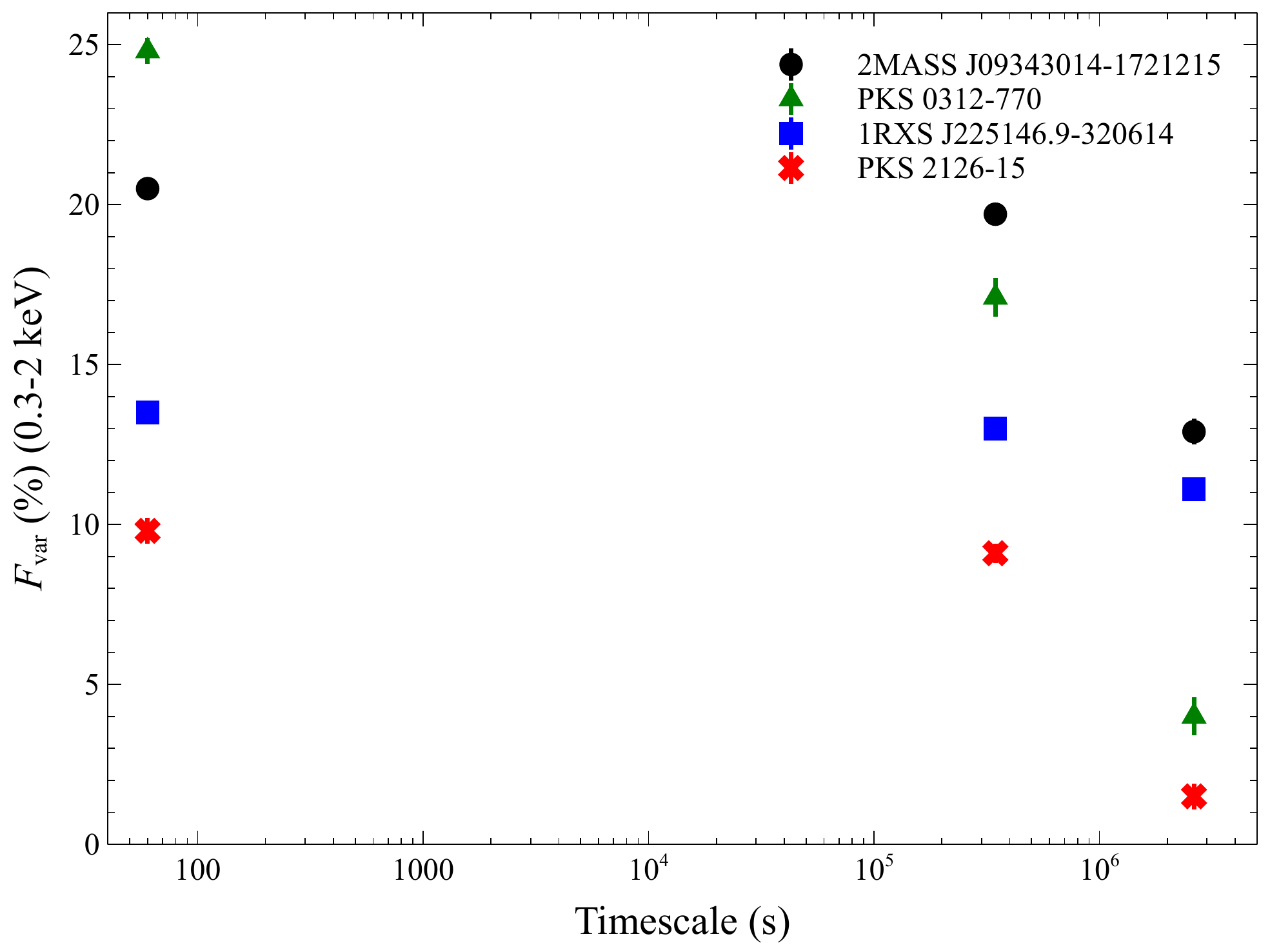}
    \caption{Fractional variability as a function of timescale. For each source, the variability amplitude decreases with increasing timescale, a trend that is not expected of blazars and other AGN.}
    \label{fig:fvarvst}
\end{figure}

\begin{figure}
	\includegraphics[width=\columnwidth]{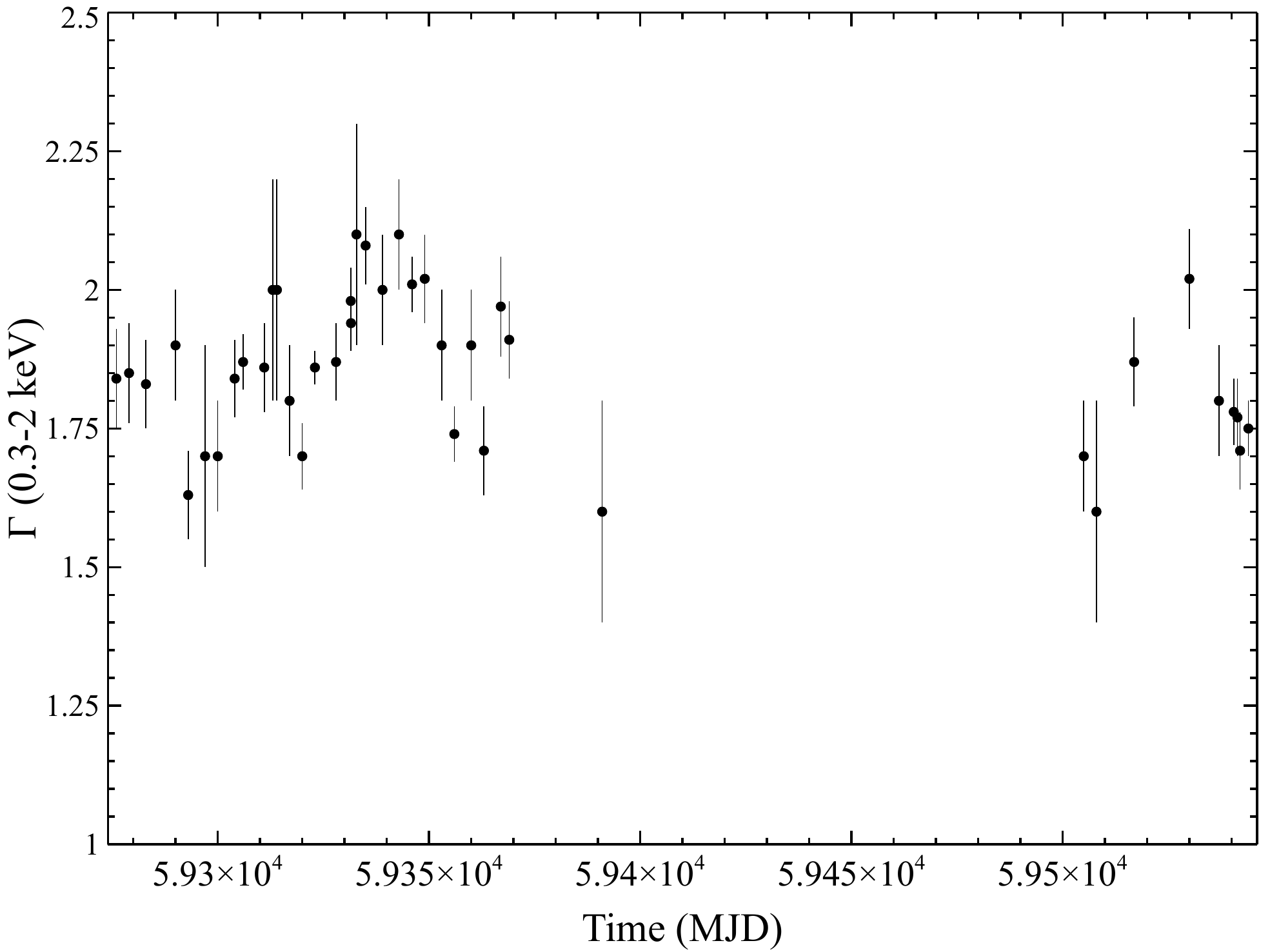}
    \caption{Photon index $\Gamma$ in the 0.3-2 keV range for 2MASS J09343014-1721215, based on spectral fits to individual observations (the gap in the middle is likely due to observing constraints). Observations with very short exposures (i.e. $\lesssim 200$s) were excluded. $\Gamma$ varies by only 7\% at the 1$\sigma$ confidence level.}
    \label{fig:2massgamma}
\end{figure}

In order to determine whether we detect variability in our sources, we fit the light curves at each timescale with a constant function, and then apply a $\chi^{2}$ test. We define a significant detection of the variability as occurring whenever $p_{\chi^{2}} \le 5$\%, where $p_{\chi^{2}}$ is the null-hypothesis probability of obtaining that value of $\chi^{2}$ if the source were in fact constant. Furthermore, we quantify the flux variability of our sources by using the methodology described in e.g. \cite{2003MNRAS.345.1271V}, which accounts for the contribution of an additional variance from measurement uncertainties. The “excess variance” has been used over the past few decades to estimate the intrinsic source variance \citep[e.g.][]{1997ApJ...476...70N,2002ApJ...568..610E}, and can be normalized to directly compare the variance between different sources. The fractional root mean square (rms) variability amplitude $F_{\rm var}$ is defined as 

%defined as the square root of the normalized excess variance \textcolor{blue}{(see Eq. \ref{eqn:fvar})}, is the quantity we will use here.

\begin{equation}
\label{eqn:fvar}
    F_{\rm var} = \sqrt{\frac{S^{2} - \overline{\sigma_{\rm err}^{2}}}{\overline{x}^{2}}}, 
\end{equation}

\noindent i.e. the square root of the normalized excess variance, and is the quantity we will use here.
%Applying this test to all 4 sources, we find that these objects show statistically significant variability.

% \begin{figure}
% 	% To include a figure from a file named example.*
% 	% Allowable file formats are eps or ps if compiling using latex
% 	% or pdf, png, jpg if compiling using pdflatex
% 	\includegraphics[width=\columnwidth]{2MASSgammavstime.pdf}
%     \caption{Photon index $\Gamma$ in the 0.3-2 keV range for 2MASS J09343014-1721215, based on spectral fits to individual observations (the gap in the middle is likely due to observing constraints). Observations with very short exposures (i.e. $\lesssim 200$s) were excluded. $\Gamma$ varies by only 7\% at the 1$\sigma$ confidence level.}
%     \label{fig:2massgamma}
% \end{figure}

Thanks to the high sensitivity of \textit{NICER}, we are in fact able to detect and quantify statistically significant variations on each of the 3 timescales we have chosen in the 0.3-2 keV band. However, we find that $F_{\rm var}$ is at most 25\% across all sources at all 3 timescales, and appears to decrease on longer timescales, implying low-amplitude variability overall and showing little variability on monthly timescales (highest $F_{\rm var}$ on monthly timescales is $12.9\pm0.4$\% across all sources, see Figure \ref{fig:fvarvst}), with 2 out of 4 sources showing $F_{\rm var}<5\%$ on monthly timescales, which is unusual and unexpected for the X-ray emission from blazars. It is possible that this, to a certain extent, may be confirming what we have observed with the lack of variability in the BAT data for these sources. While we detect statistically significant variability, this low-amplitude variability would be deeply at odds with the past general literature that shows that blazars are extremely variable objects in the X-rays. Furthermore, the behavior of the fractional variability as a function of timescale is peculiar, since for other AGN such as Seyferts, the opposite trend of variability increasing with timescale is observed \citep[e.g.][]{1999ApJ...514..682E,2002MNRAS.332..231U,2003ApJ...593...96M,2004MNRAS.348..783M,2005MNRAS.363..586U}.

While on average, the amplitude of the variability in our sources is quite low, we do find that the minute-timescale light curves exhibit occasional short bursts/flare-like events, along with periods of quiescence that last on the order of several weeks (see e.g. Figure \ref{fig:pks2126lc}, top panel (a) and bottom panel). This combination of long, low-amplitude components and short, flaring components suggests that the nature of the variability on very short timescales may be non-stationary, as opposed to the expected stationary or ``weakly" stationary variability from most other accreting objects \citep[see e.g.][]{2003MNRAS.345.1271V,2005MNRAS.359..345U}.

\subsection{Spectral analysis} \label{subsec:spec}

We fit the \textit{NICER} spectra for each source using XSPEC v12.12.0g (Arnaud 1996). Following our search for variability in the flux, we begin by searching for any signs of spectral variability from observation to observation, i.e. on $\sim$weekly timescales. In order to do this, we fit the individual observations of each source with a simple power law, using the \texttt{tbabs} model \citep{2000ApJ...542..914W} and cross-sections from \cite{1996ApJ...465..487V} to model the Galactic absorption (i.e. \texttt{tbabs*po}), setting $N_{\rm H, Gal}$ to the appropriate values from the HI4PI survey for each source \citep{2016A&A...594A.116H}. 

The energy range of the spectra used for each source in this step varies slightly, and depends on where the background starts to dominate the signal; as previously mentioned, the lowest energy where this happens, across all sources, is $\sim$2 keV, due to the generally higher background in these cases. However, some of the brighter sources, such as 1RXS J225146.9-320614 ($\sim$8 counts s$^{-1}$ in the 0.3-2 keV band), have some observations that have good-quality data out to $\sim$7 keV, so we expand the range used in fitting the individual spectra whenever appropriate, on a per source basis (the lowest energy used in all spectra is always 0.3 keV). We exclude any observations that have very low exposures (i.e. $\lesssim200$s) from the spectral variability analysis, so as to maximize the signal-to-noise and be able to constrain relevant parameters. For each of the individual observations, a simple power law modified by absorption (consistent with Galactic absorption for all but PKS 2126-15) is an acceptable fit to the data. We find that PKS 2126-15 requires absorption in excess of the Galactic absorption in order to adequately describe its individual spectra (e.g. $\Delta\chi^{2} = 132$ for one additional free parameter, for a relatively high exposure ($\sim$4 ks) observation). Therefore, for the purposes of the search for spectral variability, we include an intrinsic absorber and use the model \texttt{tbabs*ztbabs*po} for the individual observations of this source.

%The lowest energy used in all spectra is always 0.3 keV.

We find that overall, the photon index $\Gamma$ of each source does not vary by more than $\sim$15\% at the 1$\sigma$ confidence level (see Figure \ref{fig:2massgamma} for an example with 2MASS J09343014-1721215). While statistically significant ($p_{\chi^{2}}<5$\%), these variations are much smaller than is usual for blazars, and are again at odds with the literature. However, they are at least consistent with the low flux variability we find, since variations in flux are usually linked to variations in the spectrum. Due to the low spectral variability, we decide to co-add the spectra of each source in order to maximize our signal-to-noise for a more detailed spectral analysis. We also include the non-contemporaneous, time-averaged 105-month BAT spectrum for each source to obtain a joint fit in the broadband X-rays.

%We also include the time-averaged 105-month BAT spectrum for each source to obtain the full picture of the spectra in the broadband X-rays.

\begin{table*}
    \centering
    \resizebox{\textwidth}{!}{\begin{tabular}{rcccc}
         \hline \hline
    % four columns, alignment for each
		Source name & 1RXS J225146.9-320614 & 2MASS J09343014-1721215 & PKS 2126-15 & PKS 0312-770\\
		\hline
		Model & \texttt{tbabs*zlogpar} & \texttt{tbabs*po} & \texttt{tbabs*ztbabs*zlogpar} & \texttt{tbabs*po}\\
		\hline
		$N_{\rm H,Gal}$ (10$^{22}$ cm$^{-2}$)* & 0.0104 & 0.0645 & 0.0445 & 0.0783\\
		$N_{\rm H,ex}$ 10$^{22}$ (cm$^{-2}$) & ... & ... & 1.1$\pm$0.2 & ...\\
		$z$* & 0.2460 & 0.2499 & 3.2680 & 0.2230\\
		$\Gamma/\alpha$ & 1.84$\pm$0.01 & 1.81$\pm$0.02 & 1.0$\pm$0.1 &2.15$\pm$0.03 \\
		$\beta$ & 0.29$\pm$0.03 & ... & 0.16$\pm$0.06 &... \\
		$K_{\rm NICER}$ (10$^{-3}$) & 4.11$\pm$0.03 & 1.49$\pm$0.02 & 5.7$^{+0.7}_{-0.6}$ & 0.96$\pm$0.02\\
		$K_{\rm BAT}$ (10$^{-3}$) & 13$\pm$3 & 1.4$\pm$0.4 & 2.6$^{+0.7}_{-0.6}$ & 4$\pm$1\\
		\hline
		$\chi^{2}$/d.o.f. & 407/422 & 349/364 & 645/598 & 241/236\\
		\hline
    \end{tabular}}
    \caption{90\% confidence level parameters for the broaband X-ray data; all parameters other than the normalization were tied between the two instruments. $K$'s are the normalizations of the respective instruments, and $\Gamma/\alpha$ are the photon index or the slope at the pivot energy (fixed at 1 keV) in the case of the log-parabola; $\beta$ is the curvature parameter, with positive values indicating convex curvature. $N_{\rm H,Gal}$ and $z$ were frozen to the corresponding values (parameters with asterisk).}
    \label{tab:bestfit}
\end{table*}

\begin{figure*}
\centering
    \includegraphics[width=0.47\textwidth]{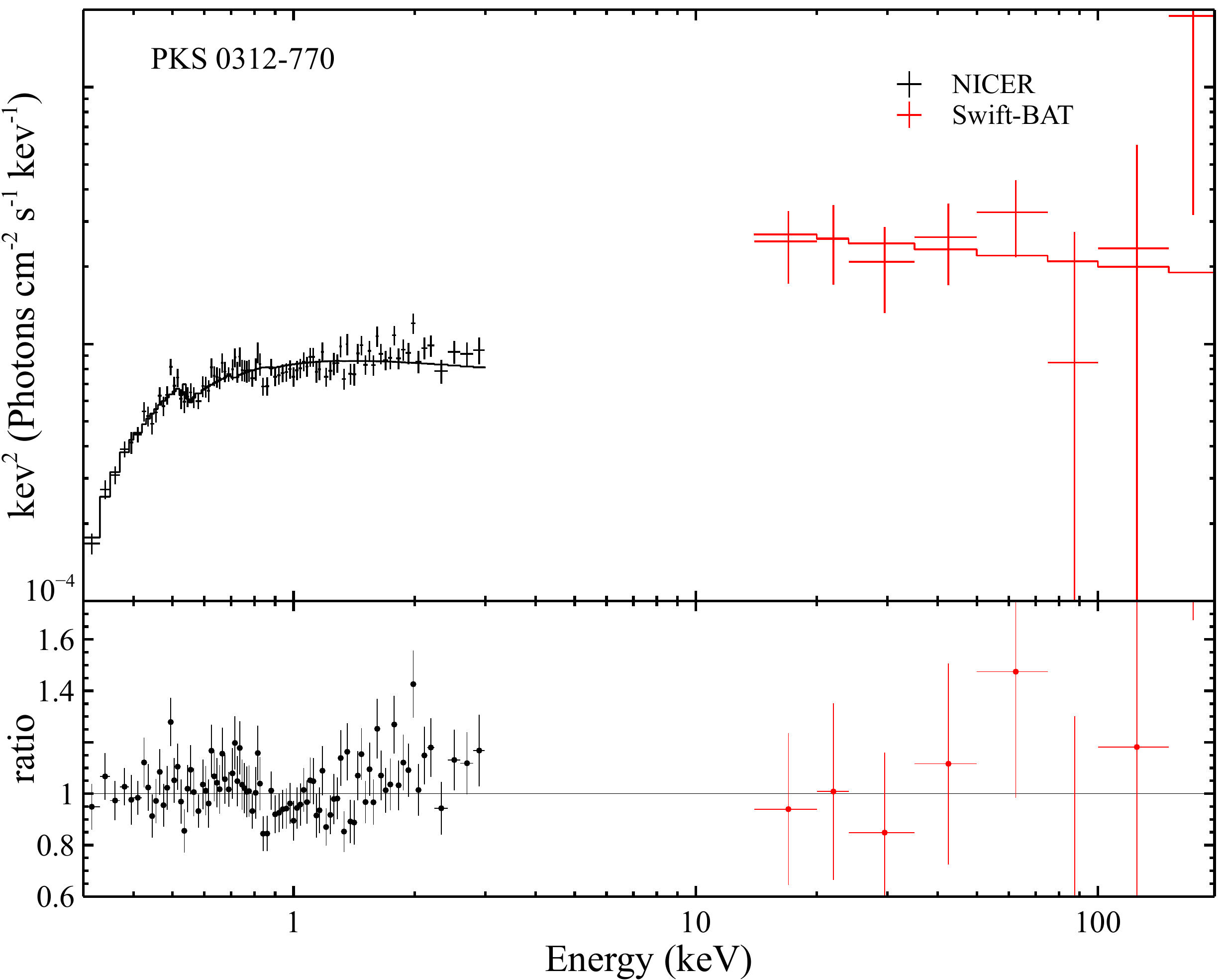}
    \includegraphics[width=0.47\textwidth]{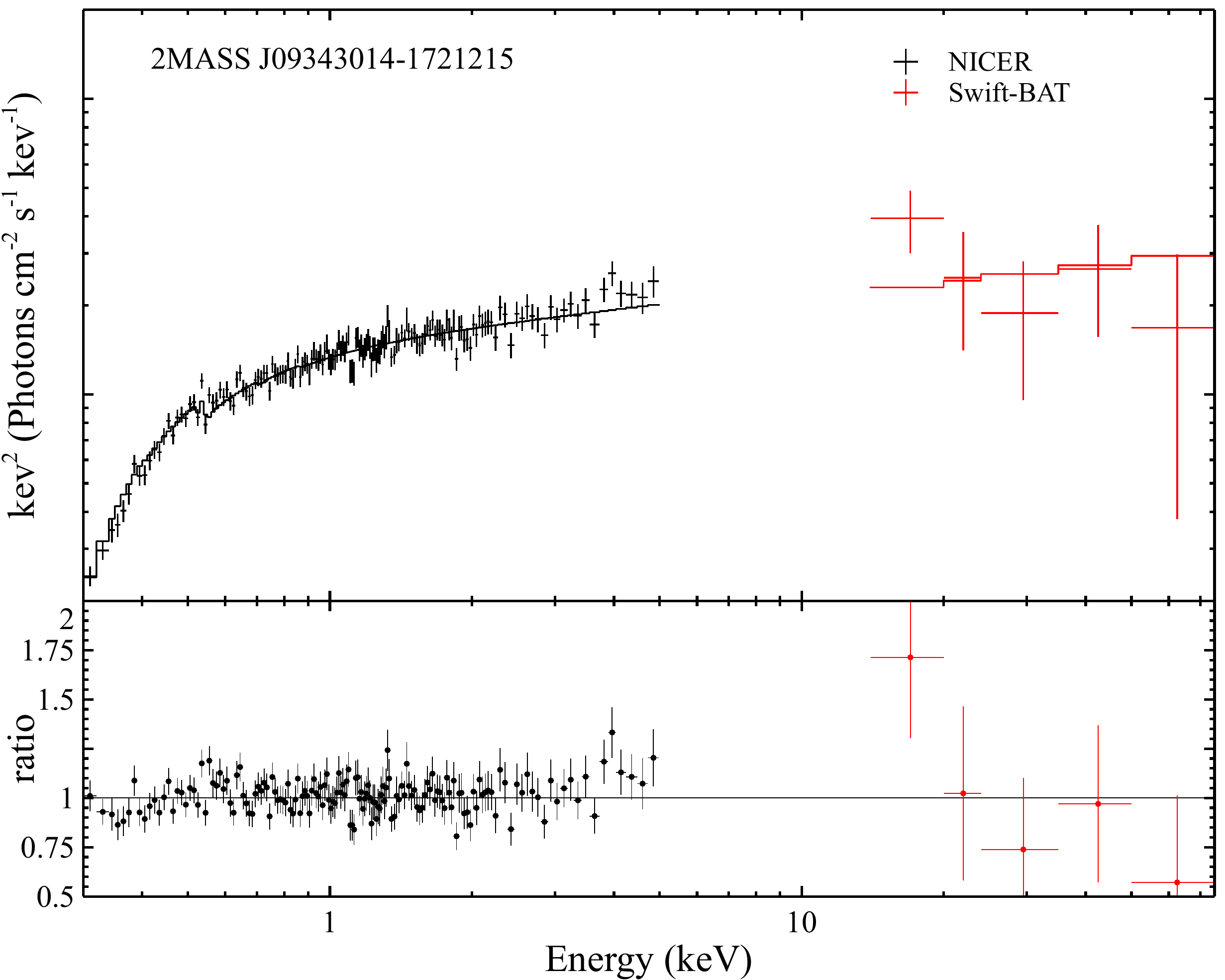}
    \includegraphics[width=0.47\textwidth]{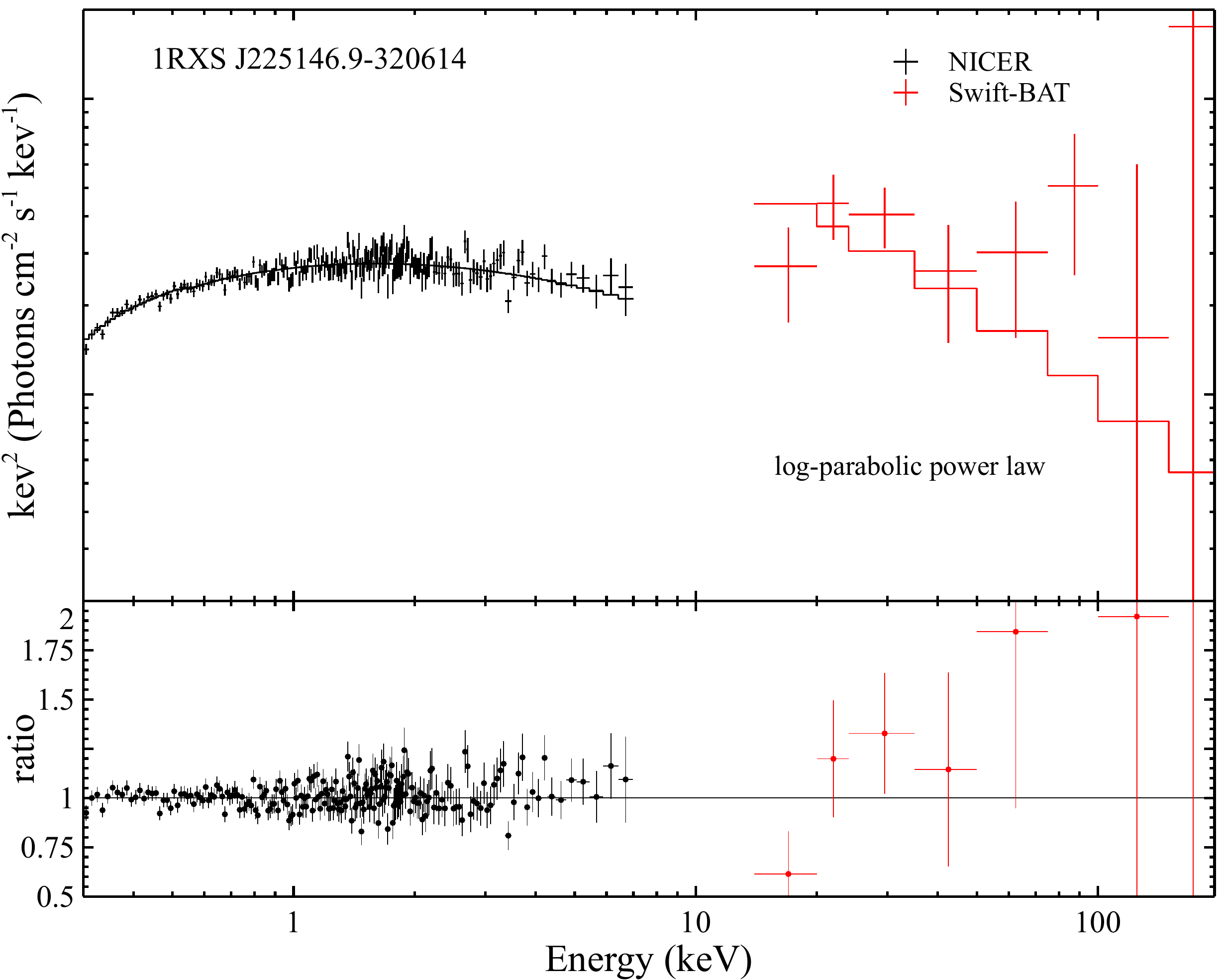}
    \includegraphics[width=0.47\textwidth]{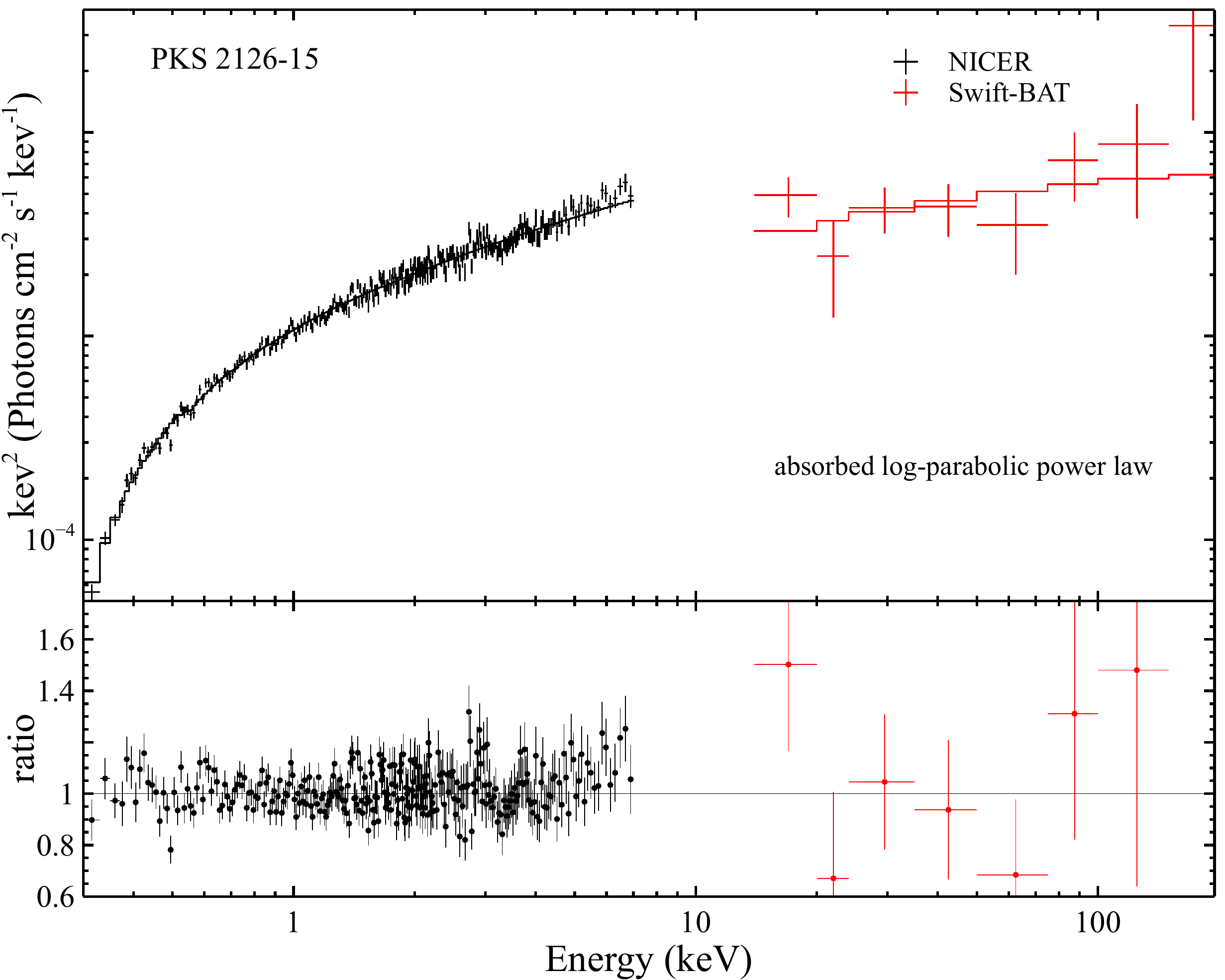}
\caption{Unfolded spectra and best-fit models for the broadband X-rays in our sample. The top two panels are fits with a simple power law.}
\label{fig:bestfits}
\end{figure*}

We start by fitting the broadband spectra with a simple power law, as we did with the individual observations, and move towards progressively more complex models when appropriate. We let the normalization vary across the two instruments to allow for any potential long-term variability, and fit with the same values for the rest of the parameters. All spectra are plotted in the rest frame of the observer. We report our results for the individual sources in the following subsections.

\subsubsection{PKS 0312-770 and 2MASS J09343014-1721215}

For PKS 0312-770 and 2MASS J09343014-1721215, the signal of the co-added spectra dominated the background out to 3 keV and 5 keV, respectively, so we use the spectra up to those energies. We also exclude the last 3 channels of the BAT data for 2MASS J09343014-1721215 due to low signal-to-noise. The data for both of these sources are well-described by a simple power law with Galactic absorption (i.e. \texttt{tbabs*po}, see Figure \ref{fig:bestfits} , top panel), with $\Gamma = 2.15\pm0.03$ for PKS 0312-770 ($\chi^{2}$/d.o.f.$=$241/236) and $\Gamma = 1.81\pm0.02$ for 2MASS J09343014-1721215 ($\chi^{2}$/d.o.f.$=$349/364). For PKS 0312-770, the normalization of the BAT data is $\sim$4 times larger than that for the \textit{NICER} data, implying some variability on very long timescales. For 2MASS J09343014-1721215, the normalization across the two instruments is practically identical (see Table \ref{tab:bestfit}), confirming a lack of variability on long timescales. 

According to \cite{2019ApJ...881..154P}, the \textit{NICER} spectrum of the BL Lac 2MASS J09343014-1721215 begins on the rising part of the synchrotron hump in the SED, and the broadband X-rays cover the very peak of the hump. In order to probe this curvature, we therefore also fit this source's spectrum with a log-parabolic power law, but we find that the curvature parameter is not well constrained and that the fit does not yield a statistically significant improvement over a simple power law.

%Maybe in discussion talk about the fact that for FSRQs expect flatter slopes, but see opposite here; reason could be that the NICER data starts on rising part of ends on very peak of synchrotron hump

\subsubsection{1RXS J225146.9-320614}

For this source, the signal of the co-added \textit{NICER} spectrum dominates the background up to $\sim$7 keV, so we use the 0.3-7 keV spectrum. When fitting with a simple power law, we find that the residuals show significant curvature. We therefore proceed to fitting with a log-parabolic power law (\texttt{tbabs*zlogpar}). This model provides the best fit to the data, improving from the simple power law fit by $\Delta\chi^{2} = 233$ for one additional free parameter, with a significance of $>99.99$\% evaluated using the $F$-test (see Figure \ref{fig:bestfits}, bottom left and Table  for fit details), with $\chi^{2}$/d.o.f.$=$407/422. We also find variability between the BAT and \textit{NICER} spectra by a factor of $\sim$3 (see normalizations in Table  ). As with 2MASS J09343014-1721215, the broadband X-ray spectrum for this source also lies on the peak of the synchrotron hump \citep{2019ApJ...881..154P}, so we associate the observed curvature with synchrotron emission intrinsic to the source, which is expected of many BL Lac objects.

\subsubsection{PKS 2126-15}

\begin{figure}
	\includegraphics[width=\columnwidth]{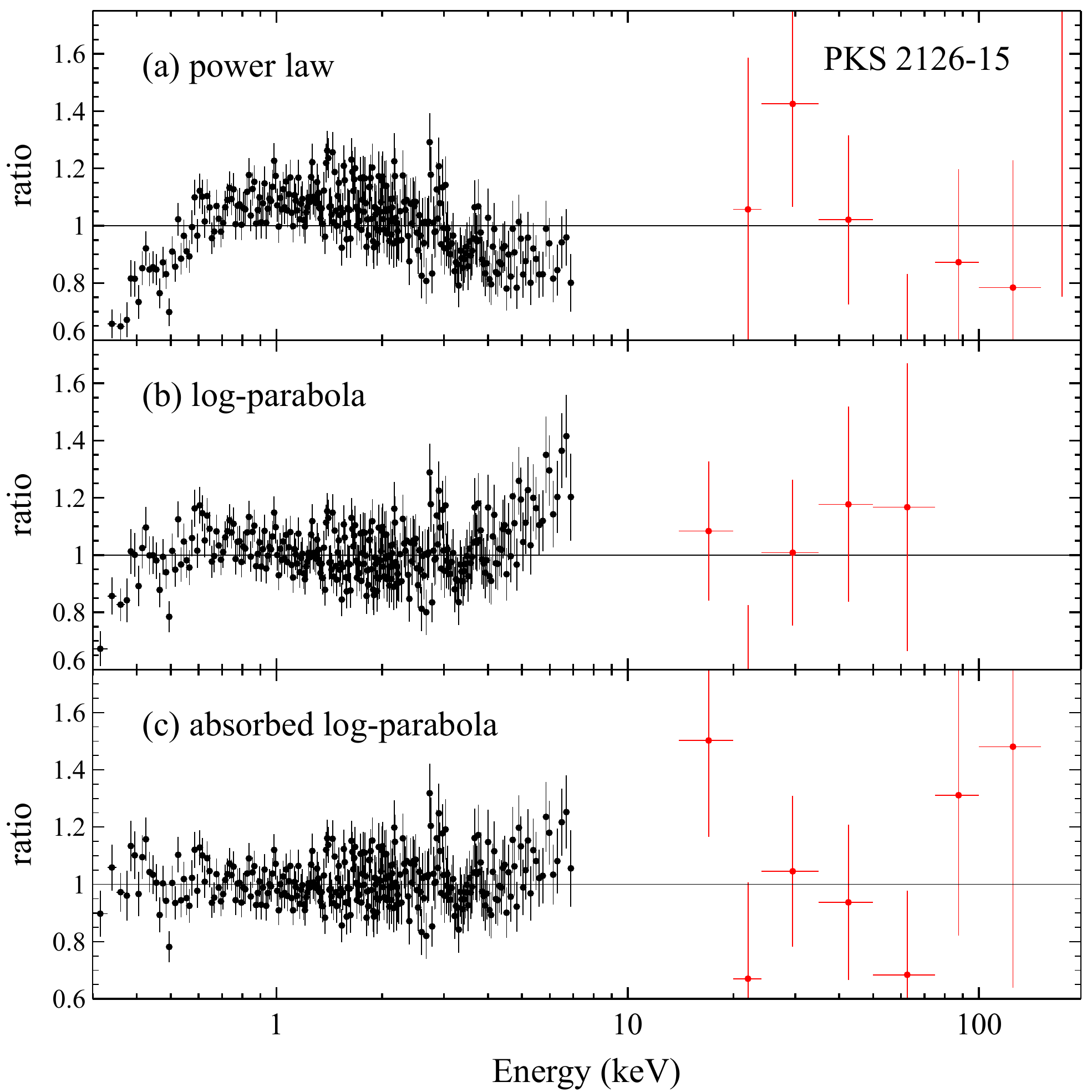}
    \caption{The ratio of \textit{NICER} and \textit{Swift}-BAT spectra for PKS 2126-15 to different models. (a) The ratio to a power law. (b) Ratio to a log-parabola. (c) Ratio to an absorbed log-parabola. The fit improves significantly with the latter (significance $>$ 99.99\%).}
    \label{fig:2126res}
\end{figure}

As with 1RXS J225146.9-320614, for PKS 2126-15 the signal dominates the background up to 7 keV. Following our usual procedure, we initially fit the co-added spectrum of PKS 2126-15 with a simple power law and once again observe curvature in the residuals. We thus proceed to fit with a log-parabolic power law. This improves the fit by $\Delta\chi^{2} = 642$ for 1 additional free parameter, at a significance of $>$99.99\%. However, this is not enough to completely describe the observed curvature, and significant negative residuals remain at the soft X-rays (see Fig. \ref{fig:2126res}). We find that an additional absorption component at the redshift of the source (\texttt{ztbabs}, assuming solar abundance), in excess of the Galactic absorption, improves the fit by $\Delta\chi^{2} = 96$ for 1 additional free parameter, at a significance of $>$99.99\%. This results in our best fit, with $\chi^{2}$/d.o.f.$=$645/598 (Figure \ref{fig:bestfits}, bottom right). We are also able to fit the spectrum with a similarly absorbed broken power law, but find that it is not statistically distinct from the absorbed log-parabola. We therefore report the fit with a log-parabolic power law, as it requires fewer free parameters (see Table \ref{tab:bestfit}).

\begin{figure}
	\includegraphics[width=\columnwidth]{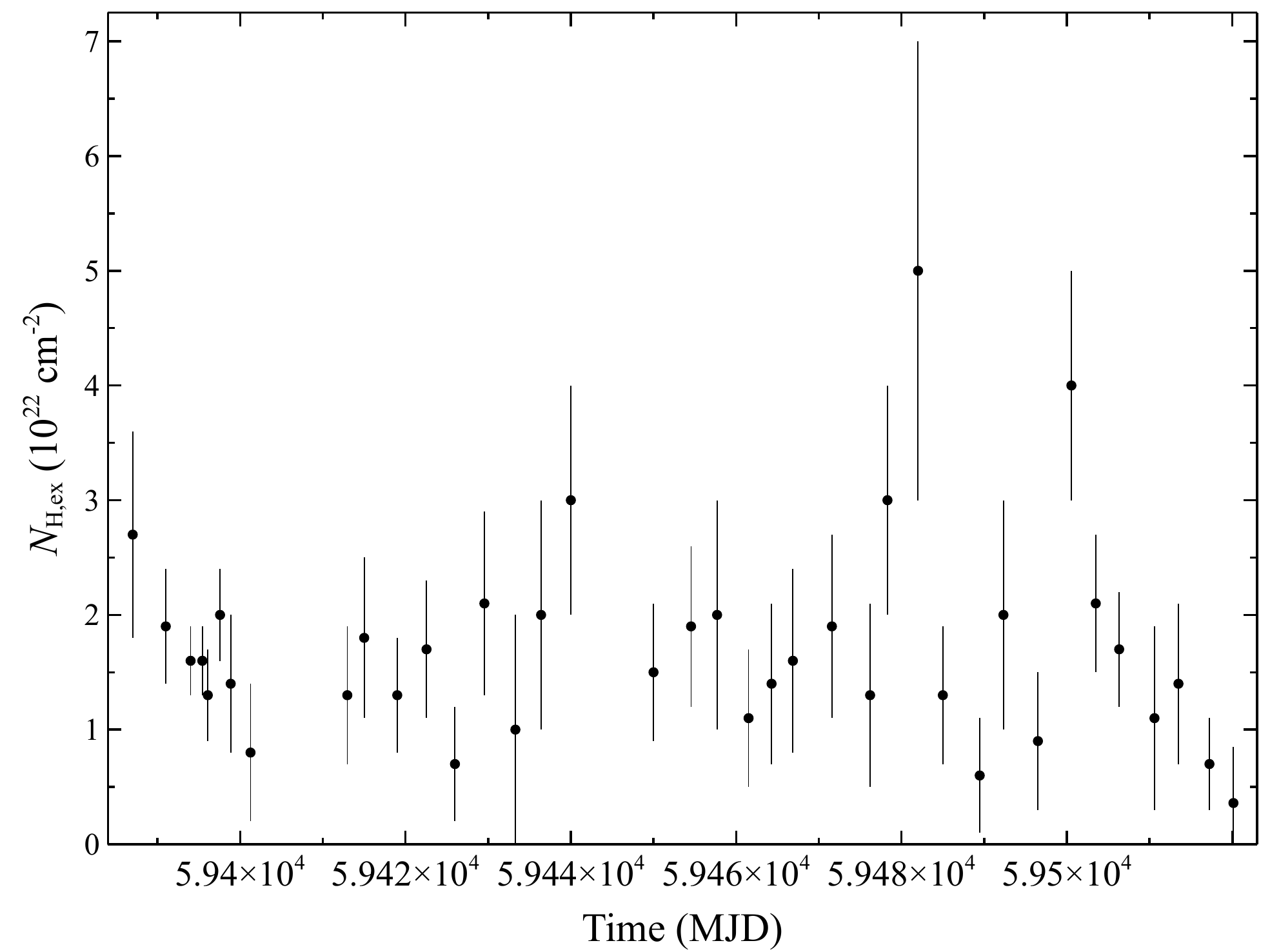}
    \caption{Time series for the column density required in excess of Galactic absorption for PKS 2126-15. A $\chi^{2}$-test indicates there is no significant variability from observation to observation.}
    \label{fig:pks2126nh}
\end{figure}

The absorption required for our best fit is significantly larger than the Galactic absorption, namely $N_{\rm H,ex}\sim10^{22}$ cm$^{-2}$ at the redshift of the source (see Table \ref{tab:bestfit}). This is unusual for blazars, as the jet is expected to remove any material from the host galaxy that may cause absorption along the line of sight. Going back to the individual observations, we find through a $\chi^{2}$ test (using the same criteria as for the flux variability analysis in Sec. 3.1) that there is no significant variability in the column density from observation to observation, with $p_{\chi^{2}}>5$\% (see Figure \ref{fig:pks2126nh}). This may hint at a possible absorption contribution from the intergalactic medium, as its column density should not change with time (see Sec. 4.2.1 for further discussion).

% (using the same criteria as for the flux variability analysis in Sec. 3.1) 

% \begin{figure}
% \gridline{\fig{0312broadbandspec.pdf}{0.5\textwidth}{a}
%           \fig{2MASSbroadbandspec.pdf}{0.5\textwidth}{b}}
% \gridline{\fig{1RXSbroadbandspec.pdf}{0.5\textwidth}{c}
%           \fig{2126broadbandspecnew.pdf}{0.5\textwidth}{d}}
% \caption{A nice inverted pyramid figure 
% consisting of six individual files}
% \end{figure}

\section{Discussion}

\subsection{Blazars not ablaze: Non-variable blazars in the broadband X-rays?}\label{sec:sec4.1}

Previous studies have shown that the extreme variability often seen in the emission of blazars may be accompanied by periods of relative quiescence. For example, \cite{2015ApJ...803...15P} and \cite{2015ApJ...807...79H} performed multi-wavelength studies of the well-studied FSRQ 3C 279, where they divided the $\gamma$-ray and X-ray light curves into flaring states and periods of low activity. Their data imply that at times, 3C 279 exhibits periods of relatively constant emission on timescales close to a year until the source undergoes a flare with extreme variability. Similarly, \cite{2022arXiv220611938C} use $\sim$3.5 years' worth of optical data from the Zwicky Transient Facility on 80 FSRQs to show that it is fairly common for transitions between active and quiescent states to occur on year-like timescales.

In another multi-wavelength study aimed at comparing the SED properties of a sample of 33 blazars, \cite{2014ApJ...789..135W} divide over 4 years of data in the optical, X-rays, and $\gamma$-rays into quiescent and active states. Much like the aforementioned studies, their data also show flaring events that are followed by low-activity epochs that last close to a year. More specifically, they find that the sources in their sample of blazars are in a $\gamma$-ray quiescent state for a significant percentage of the time, with the sources spending $<$25\% of the time in an active period, implying emission that is mostly quiescent with occasional flaring events. Studies such as \cite{2014A&A...563A..57S} and \cite{2022arXiv220611938C} report blazars that seem to not exhibit much variability for even longer periods of time in the hard X-rays and the optical band, respectively, but they exclude these objects from their analysis, as they constitute only a small percentage of their sample. For the purposes of this study, it would be ideal to compare our results to sources with similar quiescent states/behavior in the X-rays. However, similar analyses of the X-ray data of blazars that do not show much variability have not been published (at times due to poor sampling in these states, see e.g. \citealt{2013A&A...552A..11R,2015ApJ...807...79H,2021ApJ...912...54R}), with many blazar studies heavily biased towards bright targets and/or active states.   

Given that our \textit{NICER} data do not reach yearly timescales, it could very well be the case that the low-amplitude variability behavior observed in our sources is only one part of the bigger picture, and that we are observing our 4 sources in quiescent states. However, we also observe a strange trend of fractional variability amplitude decreasing with increasing timescale, which appears to be in tension with past studies that show that the stochastic variability of AGN, including blazars, is usually best described by a ``red noise" power spectrum, implying that the amplitude of the variations should increase on longer timescales (see e.g. \citealt{1999ApJ...514..682E,2002MNRAS.332..231U,2003ApJ...593...96M,2004MNRAS.348..783M,2005MNRAS.363..586U} for Seyfert-like AGN, and \citealt{2008ApJ...689...79C,2012ApJ...749..191C,2010ApJ...722..520A,2022ApJ...927..214G} for blazars). While the variability behavior of our sources seems to contradict these studies, we should stress that our data are likely too limited to constrain this type of behavior at this time. In the future, the acquisition of further data (see Section \ref{sec:sec4.1}) may allow us to compare more directly with previous studies and determine whether the nature of the variability is in fact in tension with a ``red noise" power spectrum.

%In addition?%
The \textit{NICER} data also show very low fractional variability amplitude on the same timescales as the data from the BAT catalog. As previously stated, for each of our sources, the 105-month BAT light curves show no statistically significant variability on monthly timescales, and, consistent with our fits to the broadband X-ray spectra, the \textit{NICER} and BAT bands probably originate from the same physical processes \citep[see][]{2019ApJ...881..154P}. This means that, under the assumption that the shape of the spectrum has not changed between the BAT and \textit{NICER} observations, the \textit{NICER} data could possibly serve as a proxy to the BAT data to infer that the broadband X-rays from these objects may not exhibit much variability for the duration of the BAT light curves, which would amount to almost a decade, much longer than the typical year-long duration of the quiescent periods previously cited.

As seen in Figure \ref{fig:bestfits} and Table \ref{tab:bestfit}, for three of our sources, we do observe some variability between the \textit{NICER} and BAT spectra. However, this does not necessarily contradict the trends of low variability that we have observed with the time-domain flux variability analysis. Our \textit{NICER} data are from the last calendar year, and the 105-month BAT spectra are time-averaged over a period from 2004-2013; this means that, while we may observe some variability between the different bands, it would be on very long timescales. These timescales are again much longer than the yearly timescales related to the periods of quiescence that were previously discussed. Therefore, the possibility of blazars with much longer periods of quiescence still remains.

%[\textcolor{blue}{``Theory" paragraph (or two) on cooling timescale, comparison with variability timescales, etc. here}]

In an attempt to address the very low-amplitude variability we observe, we compare our variability timescales with the cooling timescales of the emitting electrons in the jets of our sources. As previously mentioned, for the BL Lacs in our sample, the X-rays are associated with synchrotron emission \citep{2019ApJ...881..154P}, with cooling time

%to interpret our low-amplitude flux variability results

\begin{equation}
t_{\rm syn} = \frac{3m_{\rm e}c}{4\sigma_{\rm T}U_{B}\gamma\Gamma} \sim \frac{7.7\times 10^{7}}{B^{2}\gamma} \rm \ s
\end{equation}

\noindent in the observer's frame \citep[e.g.][]{1996ApJ...473..204S,Ghisellini_2013}, where $\Gamma$ is the bulk Lorentz factor, $U_{B}$ the magnetic energy density, and $\gamma$ the Lorentz factor of the relativistic electrons. Using values of $\gamma$ and the magnetic field strength $B$ from the modeling in \cite{2019ApJ...881..154P}, we find cooling times of $\sim$1 hr for 2MASS J09343014-1721215 and 1RXS J225146.9-320614. For the FSRQs in our sample, the X-ray emission ($E_{\rm x}\sim1$ keV) is caused by external Compton, with cooling time

\begin{equation}
t_{\rm EC} = \frac{3m_{\rm e}c}{4\sigma_{\rm T}u'}[E_{0}(1+z)/E_{\rm x}]^{1/2} 
\end{equation}

\noindent in the frame of the observer \citep[see e.g.][]{2013ApJ...766L..11S,2015ApJ...803...15P}, where $E_{0}$ is the characteristic energy of the seed photons, and $u'$ is the total seed photon energy density in the comoving frame, assuming mostly inverse Compton scattering of BLR and dusty torus photons, and calculated as in e.g. \cite{2009ApJ...704...38S}. We find cooling timescales on the order of $\sim$1 week for PKS 0312-770 and PKS 2126-15.

In general, the cooling times we obtain are relatively short compared to the longest timescales we probe with the \textit{NICER} and BAT data, and as we have shown, we do observe low-amplitude variability on timescales similar to some of the cooling times. Assuming a leptonic model, this could result from very short, shock-related acceleration timescales that would imply almost instantaneous acceleration, with the latter providing a near-constant source of high-energy electrons to the emitting region \citep[see e.g.][]{2009MNRAS.393.1063T}. This ``continuous" particle injection would then provide a very tight balance with the effects of energetic losses, resulting in an observed flux that is close to constant over time. Studies such as \cite{2009MNRAS.399L..59T} have in fact suggested similar scenarios to explain the apparent long-timescale stability of the TeV flux in some ``extreme" blazars. In these models, there must be very fine tuning between injection and loss processes to produce the observed roughly constant fluxes, a strong constraint on blazar models.

%Studies such as Tavecchio et al. (2009) have suggested similar scenarios to explain the stability of the TeV flux in some extreme blazars  

%to the "radiating" region?

Of course, we also observe variability with slightly higher $F_{\rm var}$ on timescales of a minute. Upon close inspection of the short-timescale light curves, we find that this is likely due to the fact that the variability is characterized by longer periods of relative quiescence with occasional events that are flare-like in nature and which last several minutes. While many accreting objects are believed to show variability behavior that is near time-stationary (i.e., the statistical moments of the underlying process remain fairly constant over time), non-stationary time series involving flares and quiescent periods have been observed in the X-rays before \citep[see e.g.][]{1997ApJ...481L..15L,2019MNRAS.482.2088A}, as well as in the aforementioned studies with Fermi-detected sources, although as previously noted, the studies on the Fermi sources correspond to baselines much longer than the one currently available for our data.

%for 3 sources

%Of course, we also observe variability with slightly higher amplitude on timescales of a minute, 

The very short-timescale variability addressed in this study occurs on timescales that are much shorter than the cooling timescales for our sources. This suggests that the variability on these timescales is likely due to other factors that are not related to energetic losses. Generally, blazar emission models assume a single quasi-spherical emission region, however it could be the case that the jets have many localized, X-ray emitting regions that contribute to this short-timescale variability, but whose effects on the emission are smoothed out when observed on longer timescales. \cite{2009MNRAS.393L..16G} proposed a multi-region scenario to explain fast, $\sim$minute timescale variability in 2 BL Lacs, with the variability potentially being caused by ultra-relativistic particles continuously flowing along magnetic field lines through magneto-centrifugal acceleration of the particle beams. They proposed this scenario in the context of ultra-fast flaring events, of which we do find some evidence in the data.

%for 3 of our sources

%Given the trend of lower variability amplitude with increasing timescale, we might infer that low-amplitude variability exists on most timescales for these sources.  }

%compared to the longest timescales probed by the \textit{NICER} and BAT data. 

It is important to note that in general, it would be premature to draw strong conclusions from these results given our small sample size and the fact that the \textit{NICER} data only go out to monthly timescales. To that end, we have a new follow-up, multi-cycle \textit{NICER} campaign currently underway that increases the sample size by 50\% and our temporal baseline by a factor of 3. This campaign will allow us to probe the variability of our sources up to timescales of a year, with the main objectives of characterizing the long-term variability, searching for potential yearly flares, and obtaining high-quality time-averaged spectra (Mundo et al. 2023, in preparation).

\subsection{Interpreting the Broadband X-ray spectra}

The X-ray spectra of blazars can usually be described by a simple power law or some form of curved continuum. Blazar emission models can in fact predict curvature in the emitted spectra that is linked to the shape of the emitting particle energy distribution \citep[e.g.][]{2007ApJ...665..980T,2009ApJ...704...38S,2009MNRAS.397..985G,2015MNRAS.448.1060G}. In other, more extreme cases, blazars may have their synchrotron peak located at frequencies $>10^{17}$ Hz, implying that the curvature and the broadband X-rays are produced exclusively by synchrotron emission \citep[e.g.][]{2019ApJ...881..154P}. Therefore, in many cases a curved spectrum is seen as a quality intrinsic to the physical processes that are at the root of blazar emission.

%which at times can itself exhibit curvature due to phenomena such as inefficient cooling of low-energy electrons and photon starving (see e.g. Gianni et al. 2011 and appendices in Arcodia et al. 2018).

We observe these spectral characteristics with our sample, with two blazars exhibiting power-law behavior and the remaining two well-described by models that invoke curvature. In particular, 1RXS J225146.9-320614 requires a log-parabolic power law as the best fit for its broadband X-ray spectrum, and we attribute this curvature to synchrotron emission, as according to \cite{2019ApJ...881..154P} it is a high-frequency peaked blazar, with the broadband X-rays falling on the synchrotron peak. The latter is also true of 2MASS J09343014-1721215, but we find that a log-parabola does not significantly alter the fit, and that the low curvature parameter is not very well-constrained. Additional data from our previously mentioned follow-up campaign would increase the quality of the spectrum and may reveal the synchrotron peak more clearly.

PKS 2126-15 also shows curvature in the form of a log-parabolic continuum that is likely intrinsic. However, this source is an FSRQ, with the broadband X-rays lying on the rising part of the inverse Compton hump and likely produced by EC effects \citep{2019ApJ...881..154P}. Therefore, the observed curvature is likely more directly connected to the shape of the energy distribution of the electrons that are upscattering the external photon fields. However, in order to fully describe the curvature, our fit also requires significant absorption at the soft X-rays in excess of the Galactic absorption, which is unusual for a blazar.

\subsubsection{The curious case of PKS 2126-15: An absorbed FSRQ?}
%Spectral hardening in the soft X-rays has been observed many times in distant AGN (see e.g. ). However, it is not always clear if this is due to 
Generally, blazars are not expected to show significant X-ray absorption in their spectra. This is because the kpc-scale jet points along our line of sight, and thus likely sweeps any potential X-ray absorption component from the host. In particular, FSRQs tend to be the most luminous, powerful blazars \citep[e.g.][]{1998MNRAS.299..433F,2017MNRAS.469..255G}, and are therefore more effective in removing host absorbers in the vicinity, further decreasing the likelihood of any type of contribution from the host galaxy to their X-ray spectra.

Despite this, over the past two decades, several blazars have shown hardening at the soft X-rays that can in fact be modeled by absorption \citep[e.g.][]{1997ApJ...478..492C,2000ApJ...543..535T,2001MNRAS.323..373F,2001MNRAS.324..628F,2004MNRAS.350..207W,2004MNRAS.350L..67W,2006MNRAS.368..844W,2005MNRAS.364..195P,2004AJ....127....1G,2006AJ....131...55G,2007ApJ...669..884S,2013ApJ...774...29E}. In these studies, the absorption was usually described by a neutral absorber intrinsic to the host galaxy. However, this is inconsistent with the significantly lower absorption seen at optical and UV wavelengths (see discussions in e.g. \citealt{1997ApJ...478..492C,2001MNRAS.323..373F,2004MNRAS.350..207W,2004MNRAS.350L..67W,2005MNRAS.364..195P}). As a result, several studies thereafter preferred explaining the curvature with intrinsic spectral breaks \citep[e.g.][]{2007ApJ...665..980T}.

As shown in \cite{2013ApJ...774...29E}, high redshift blazars, most of which are FSRQs, are often absorbed. Two recent studies by \cite{2018A&A...616A.170A} and \cite{2021MNRAS.508.1701D} find that absorption in excess to the Galactic absorption, as opposed to intrinsic breaks, is preferred to fully explain the hardening in their respective samples of FSRQs. They successfully describe the excess X-ray absorption as occurring due to the highly ionized ``warm-hot" intergalactic medium (WHIM), which would also account for the lack of absorption seen in the optical/UV. In their analyses, they directly model the WHIM and perform simultaneous fits with their sources to measure its properties. They both calculate mean hydrogen densities of $n_{\rm 0}\sim10^{-7}$ cm$^{-3}$ (at $z=0$) in the WHIM and temperatures of log($T/K$)$\sim$6, which are consistent with the quantities expected of such a medium (see simulations in e.g. \citealt{1999ApJ...514....1C,2006ApJ...650..560C,2007MNRAS.374..427D,2015MNRAS.446..521S,2019MNRAS.486.3766M}. \cite{2021MNRAS.508.1701D} also combine their blazar data with that of gamma-ray bursts to find that their results agree over a wide range in redshift, and show that the WHIM column density is not related to variations in the flux or spectra of their sources. 

The high signal to noise of the \textit{NICER} data, combined with the broad bandwidth of including the BAT data, allow us to both measure curvature in the continuum and require ``excess" absorption in PKS 2126-15. In other words, even with a more sophisticated treatment of the continuum of PKS 2126-15 (i.e. accounting for intrinsic curvature), we still require absorption that is consistent with the results of \cite{2018A&A...616A.170A} and \cite{2021MNRAS.508.1701D}. While such rigorous studies are beyond the scope of this paper, excess absorption in the WHIM could be one way to interpret what we observe for PKS 2126-15 in the soft X-rays, given that it is a high-$z$ FSRQ. Absorption in the intergalactic medium would not change over time, since it should not depend on the source's environment, and we observe a lack of variability in the column density at the redshift of PKS 2126-15, over many observations. While this of course does not present definitive evidence, it might point towards a scenario involving the WHIM. In addition, since the WHIM is expected to be diffuse and smeared over redshift, it is possible for its signature to also appear at or near the redshift of the source. In the future, and especially with the arrival of the data from our new multi-cycle \textit{NICER} campaign, we hope to robustly probe this scenario with more physically motivated models.      

%are thus far the only two studies that successfully attribute the spectral hardening to absorption in the ``warm-hot" intergalactic medium   

%(e.g. Arcodia et al. 2018, Dalton et al. 2021).    

%Therefore, one way to interpret the significant absorption that we observe in the soft X-rays is to infer that there is a possible contribution from the IGM. Absorption in the IGM would not change over time, given that it should not depend on the source's environment. In addition, the IGM is diffuse and smeared over redshift, meaning that it is possible for its signature to also appear at or near the redshift of the source.

%obtained in quiescent and active states in the optical, X-rays, and $\gamma$-rays.Williamson et al. (2017) also conducted a multi-wavelength study on a sample 33 blazars that divides the data into quiescent and active states, with the intent of comparing the spectral indices obtained in quiescent and active states in the optical, X-rays, and $\gamma$-rays.     

\section{Conclusions}

We have presented X-ray spectral and time-domain variability analyses of 4 ``quiescent" blazars from the \textit{Swift}-BAT 105-month catalog using \textit{NICER} data from a recent 5-month long campaign, as well as archival BAT data. Our main results are as follows:

\begin{enumerate}
    \item We detect statistically significant, but very low-amplitude ($F_{\rm var}<25\%$) variations in the flux of our sources on three distinct timescales, which is at odds with the expected high-amplitude variability of blazars.
    \item For each source, the fractional variability decreases with increasing timescale, in general showing low variability on monthly timescales ($F_{\rm var}\lesssim13\%$), which is in tension with the ``red noise" variability usually observed in AGN. This could imply a very constrained scenario where near-continuous particle injection balances the effects from energetic losses.
    \item The minute-timescale variability appears to be characterized by non-stationary behavior involving long periods of quiescence with occasional bursts, with the latter possibly caused by processes similar to those that lead to ultra-fast flares.
    \item As is customary with blazars, we are able to fit the broadband X-ray spectra with different power law models, with two sources that are well-described with a simple power law and two sources that require curvature in the form of at least a log-parabola. 
    \item For PKS 2126-15, a high-$z$ FSRQ, we require a column density significantly higher than that for Galactic absorption ($N_{\rm H,ex}\sim10^{22}$ cm$^{-2}$). We posit that this may be due to a possible absorption contribution from the warm-hot intergalactic medium.  
\end{enumerate}

\section*{Acknowledgements}

SM and RM acknowledge support from NASA grant 80NSSC21K1995. 

% %%%%%%%%%%%%%%%%%%%%%%%%%%%%%%%%%%%%%%%%%%%%%%%%%%
\section*{Data Availability}

Supplementary data such as additional figures and the entirety of Table \ref{tab:obs} are available in the electronic version or upon request (contact Sergio A. Mundo). The \textit{NICER} data used for the blazar spectral and time-domain variability analyses are available online at \href{https://heasarc.gsfc.nasa.gov/cgi-bin/W3Browse/w3browse.pl}{https://heasarc.gsfc.nasa.gov/cgi-bin/W3Browse/w3browse.pl}. Archival \textit{Swift}-BAT data for the blazars are available at \href{https://swift.gsfc.nasa.gov/results/bs105mon/}{https://swift.gsfc.nasa.gov/results/bs105mon/}.



\bibliographystyle{mnras}
\bibliography{example} % if your bibtex file is called example.bib

%\begin{thebibliography}{99}
%\end{thebibliography}

%%%%%%%%%%%%%%%%%%%%%%%%%%%%%%%%%%%%%%%%%%%%%%%%%%

%%%%%%%%%%%%%%%%% APPENDICES %%%%%%%%%%%%%%%%%%%%%

%\appendix



%%%%%%%%%%%%%%%%%%%%%%%%%%%%%%%%%%%%%%%%%%%%%%%%%%

% Don't change these lines
\bsp	% typesetting comment
\label{lastpage}
\end{document}